\documentclass[prl,twocolumn,amsmath,amssymb]{revtex4}
\usepackage{graphicx}
\usepackage{dcolumn}
\usepackage{bm}

\begin{document}



\title{Infrared spectroscopy of electronic bands in bilayer graphene}

\author{A. B. Kuzmenko, E. van Heumen, D. van der Marel}
\affiliation{D\'{e}partment de Physique de la Mati\`{e}re
Condens\'{e}e, Universit\'{e} de Gen\`{e}ve, CH-1211 Gen\`{e}ve
4, Switzerland}
\author{P. Lerch}
\affiliation{Paul Scherrer Institute, Villigen 5232,
Switzerland}
\author{P. Blake, K. S. Novoselov, A. K. Geim}
\affiliation{Manchester Centre for Mesoscience and
Nanotechnology, University of Manchester, Manchester M13 9PL,
UK }

\begin{abstract}
We present infrared spectra (0.1-1 eV) of electrostatically
gated bilayer graphene as a function of doping and compare it
with tight binding calculations. All major spectral features
corresponding to the expected interband transitions are
identified in the spectra: a strong peak due to transitions
between parallel split-off bands and two onset-like features
due to transitions between valence and conduction bands. A
strong gate voltage dependence of these structures and a
significant electron-hole asymmetry is observed that we use to
extract several band parameters. Surprisingly, the structures
related to the gate-induced bandgap are much less pronounced in
the experiment than predicted by the tight binding model.
\end{abstract}

\maketitle

Since the first successful attempt to isolate graphene
\cite{NovoselovScience04}, this two-dimensional material
remains in the focus of active research motivated by a unique
combination of electronic properties and a promising potential
for applications \cite{GeimNovoselovNatMat07}. Its infrared
response, like many other transport and spectral properties, is
notably distinct from the one of conventional metals and
semiconductors. For example, the optical conductance $\mbox{Re
}G(\omega)$ of monolayer graphene, which describes the photon
absorption by a continuum of electronic transitions between the
hole and electron conical bands, remains constant in a broad
range of photon energies and equal to $G_{0} = e^2/4\hbar$
\cite{AndoJPSJ02,NairScience08,LiNP08}. Quite remarkably, the
optical transmittance of single carbon layer in this range
depends solely  on the fine structure constant
\cite{KuzmenkoPRL08,NairScience08}. In bilayer graphene, where
the interlayer electron hopping results in two extra electron
and hole bands separated from the main bands by about 0.4 eV,
one expects to see a set of intense and strongly doping
dependent infrared structures
\cite{NilssonPRL06,AbergelFalkoPRB07,NicolCarbottePRB08}
sensitive to various band details and quasiparticle scattering
rates. This makes infrared spectroscopy a powerful probe of the
low-energy electronic dispersion in graphene, especially in
combination with a possibility to electrostatically control the
doping level \cite{JiangPRL07,LiNP08,WangScience08}. Here we
present infrared spectra of bilayer graphene crystals in a
broad doping range, which allows us to observe new features, in
particular a significant electron-hole asymmetry. By comparing
data with the tight binding Slonczewski-Weiss-McClure (SWMcC)
model \cite{SWMcC} we identify interband transitions and
determine some band parameters.

Bilayer graphene is considered to be particularly important for
electronics applications by virtue of a bandgap that opens when
there is difference between the electrostatic potential of the
two layers
\cite{McCannFalkoPRL06,GuineaPRB06,McCannPRB06,OhtaScience06,CastroPRL07,OostingaNatMat08}.
Angle-resolved photoemission (ARPES) measurements
indicate such a gap in potassium doped bilayer graphene
epitaxially grown on SiC \cite{OhtaScience06}. Although
transport experiments \cite{CastroPRL07,OostingaNatMat08}
demonstrate that a bandgap also opens in gate-tunable bilayer
graphene flakes, no spectroscopic information about the size of
the gate-induced gap is currently available. The analysis of infrared data
allows us to get further insight into this issue.

\begin{figure}[thb]
   \centerline{\includegraphics[width=8cm,clip=true]{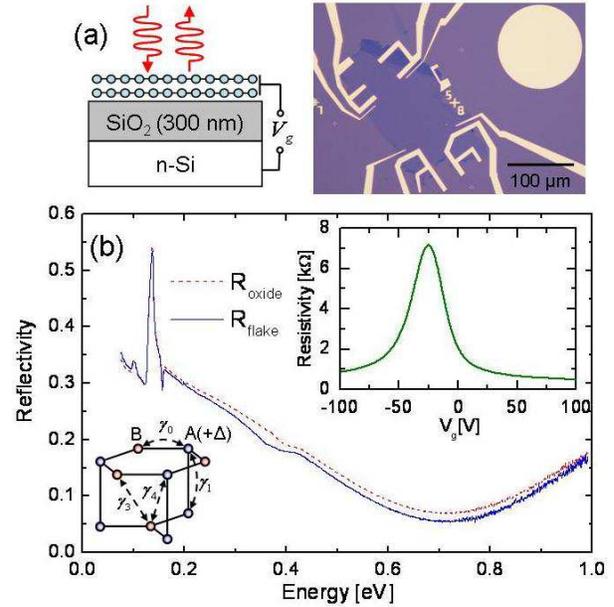}}
   \caption{(a) Schematic view and a micrograph of the used bilayer graphene device.
   (b) Infrared reflectance of graphene flake (blue solid line) and of
   bare substrate (red dotted line) (taken at $T$ = 10 K and $V_{g}$ = +100 V).
   Left inset: Bernal stacking of bilayer graphene and relevant hopping terms, right inset:
   resistivity at 10 K as a function of the gate voltage.}
   \label{Fig1}
\end{figure}

The sample used in this study is a large ($\sim$100 $\mu$m)
bilayer graphene flake (Graphene Industries Ltd.) on top of an
n-doped Si substrate covered with a 300 nm layer of SiO$_{2}$
(Fig.\ref{Fig1}(a)). A field-effect device configuration
allowed us to simultaneously measure the DC resistivity and
infrared reflectance as a function of the applied gate voltage
$V_{g}$. Optical spectra in the photon energy range 0.1-1 eV
were collected at $\approx$ 10 K with an infrared microscope
(Bruker Hyperion 2000) focussing the beam on a spot of about 30
microns in diameter. The absolute reflectance of graphene,
$R_{\mbox{\scriptsize flake}}$, and of the bare substrate,
$R_{\mbox{\scriptsize oxide}}$, (Fig.\ref{Fig1}(b)) were
obtained by using a circle of gold deposited close to the
sample as a reference mirror. The bare substrate spectrum
features intense optical phonon modes in SiO$_{2}$ below 0.15
eV and a dip at 0.7 eV due to the Fabry-Perot effect in the
SiO$_{2}$ layer. The change of the absolute reflectivity
introduced by graphene $\Delta R = R_{\mbox{\scriptsize flake}}
- R_{\mbox{\scriptsize oxide}}$ is small but reproducibly
measurable as we checked on a second sample. By taking
difference spectra, we largely cancel spurious optical effects
such as a weak 0.4 eV absorption band due to some frozen water.
The resistivity maximum that corresponds to zero doping
(Fig.\ref{Fig1}(b), inset) is found to be at $V_{g0}$ = -25 V
instead of 0 V, which we attribute to a charging effect by
contaminant molecules.

\begin{figure}[thb]
   \centerline{\includegraphics[width=6cm,clip=true]{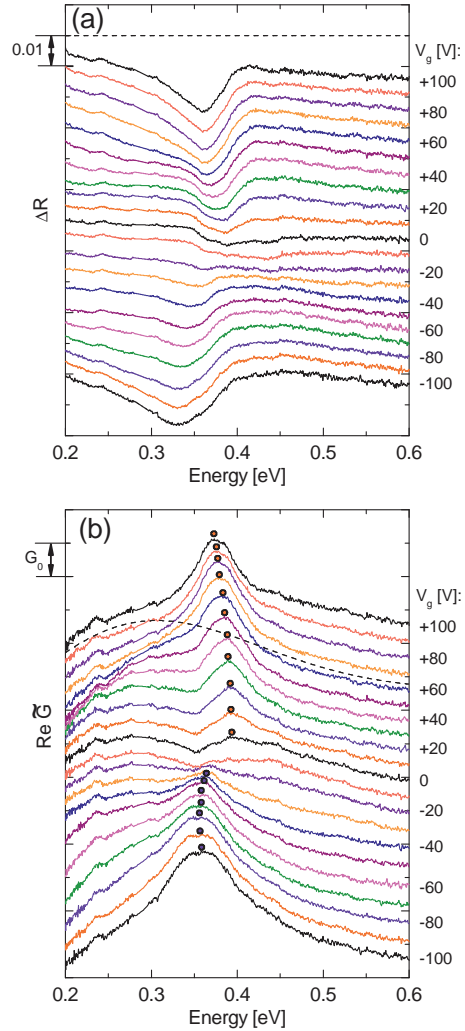}}
   \caption{(a) Mid-infrared spectra of $\Delta R$ at $T \approx$ 10 K as a function of the gate voltage $V_{g}$.
   The curves are separated by 0.005; the dashed line is the zero level for the +100 V curve.
   (b) Real part of the infrared sheet conductance of bilayer graphene $\tilde{G}(\omega)$,
   derived from the reflectance curves (Fig.2a) using a Kramers-Kronig inversion.
   The curves are separated by $0.5 G_{0}$. Note that $\tilde{G}(\omega)$ possibly differs
   from the true conductance $G(\omega)$ by a spectrally featureless gate-independent background, as explained in the text.
   The dashed line is the correction (shown relative to the +100 V  spectrum) used to generate Fig. \ref{Fig3}b.
   }
   \label{Fig2}
\end{figure}

The curves of $\Delta R(\omega)$ between 0.2 and 0.6 eV are
shown in Fig.\ref{Fig2}(a) as a function of the gate voltage
from -100 V to +100 V. The spectra in this region are very
sensitive to the gate voltage and show a significant asymmetry
between the electron ($V_{g} > V_{g0}$) and the hole ($V_{g} <
V_{g0}$) doping. Since it is more convenient to discuss the
data in terms of the real part of the optical conductance
$G(\omega)$, we extracted this quantity by a Kramers-Kronig
(KK) constrained inversion \cite{KuzmenkoRSI05} of the raw
reflectivity data. Due to a sensitivity of the inversion
procedure to the systematic uncertainty ($\sim $ 0.005) of
$\Delta R$ and to the data extrapolations beyond the
experimental spectral range (we used graphite optical data
\cite{KuzmenkoPRL08} as the most reasonable extrapolation) the
inverted function $\mbox{Re } \tilde{G}(\omega)$ is likely to
contain a spectrally smooth background as compared to $\mbox{Re
} G(\omega)$. Although this background does not allow us to
determine accurately the absolute conductance, it affects the
positions of spectral structures and their doping dependence to
a much lesser extent.

\begin{figure}[thb]
   \centerline{\includegraphics[width=\linewidth,clip=true]{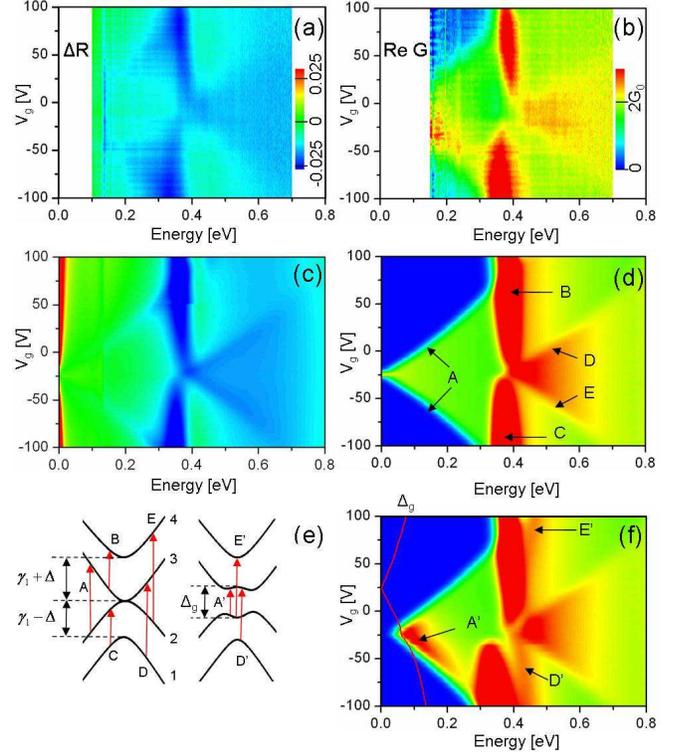}}
   \caption{(a) and (b): color plots of the raw $\Delta R(\omega)$ and the derived $\mbox{Re }G(\omega)$ spectra as a function of $\omega$ and
   $V_{g}$. (c) and (d): $\Delta R$ and $\mbox{Re }G(\omega)$ calculated
   using the tight-binding model assuming that the bandgap is
   zero. (e): the four bands of bilayer graphene in the absence
   (left) and in the presence (right) of the bandgap, with the interband transitions shown with arrows.
   (f): $\mbox{Re }G(\omega)$ calculated assuming that the
   bandgap $\Delta_{g}$ is present as given by the red solid curve (Ref.\cite{CastroPRL07}).
   }
   \label{Fig3}
\end{figure}

The spectra of $\mbox{Re }\tilde{G}(\omega)$
(Fig.\ref{Fig2}(b)) reveal a prominent peak centered between
0.35 and 0.4 eV, whose intensity increases with the absolute
value of the gate voltage and vanishes as $V_{g}$ approaches
$V_{g0}$. Based on previous theoretical works
\cite{NilssonPRL06,AbergelFalkoPRB07,NicolCarbottePRB08} as
well as on the calculations described below we assign this peak
to a transition between the hole bands 1 and 2 (marked as C in
Fig.\ref{Fig3}(e)) for $V_{g} < V_{g0}$ and to the one between
the electron bands 3 and 4 (marked as B) for $V_{g} > V_{g0}$.
The doping induced shift of the Fermi level away from the Dirac
point expands the momentum space, where this transition is
allowed by the electronic occupation of the initial and the
final states, and therefore increases the infrared intensity of
the peak.

The energy of this peak is given by the band separation and is
close to the interlayer vertical hopping parameter $\gamma_{1}$
(shown in the inset of Fig.\ref{Fig1}(b)). In the case of
precisely symmetric electron and hole bands, one would expect
the same peak position for the positive and negative gate
voltages. However, the data reveal a clear asymmetry: at
positive voltages the maximum (marked with red circles in
Fig.\ref{Fig2}(b)) is higher in energy and shows a much
stronger dependence on $V_{g}$ than at negative voltages (blue
circles). As was pointed out in Ref.\onlinecite{LiCM08}, the
energy of the peak on the electron and hole side taken close to
the charge neutral point ($V_{g0}$ = -25 V in our case) is
equal to $\gamma_{1}+\Delta$ and $\gamma_{1}-\Delta$
respectively, where the parameter $\Delta$ is the potential
difference between carbon sites A and B. These values in our
case are 0.393 $\pm$ 0.005 eV and 0.363 $\pm$ 0.005, which
yields $\gamma_{1} = 0.378 \pm 0.005 $ eV and $\Delta = 0.015
\pm 0.005 $ eV. The value of $\gamma_{1}$ is very close to
0.377 eV found in graphite \cite{ChungReview}. However, it is
somewhat smaller than 0.404 eV reported in
Ref.\onlinecite{LiCM08} for bilayer graphene flake. This
suggests that the interlayer distance, to which $\gamma_{1}$ is
the most sensitive, may change from sample to sample. As far as
$\Delta$ is concerned, there is much less agreement on the
value of this parameter in graphite in the literature. While
the magnetoreflection and de Haas - van Alphen measurements
suggest that $\Delta$ is -0.008 eV (see
Ref.\onlinecite{ChungReview} and references therein), infrared
data \cite{GuizzettiPRL73,KuzmenkoPRL08} give a value of +0.04
eV. Our value agrees in sign with the infrared-based estimate
in graphite, but is about 2-3 times smaller. This difference
can be understood using electrostatics arguments. In Bernal
stacked graphite, each carbon layer is symmetrically surrounded
by {\em two} other layers, in contrast to bilayer graphene.
Therefore one may expect the difference between the (screened)
Coulomb potential on sites A and B induced by charges on other
layers to be larger in graphite.

In order to get further insight, we compare the experimental
data with calculations based on the tight binding SWMcC model
that proved to be very successful in
graphite\cite{SWMcC,PartoensPRB06,KuzmenkoPRL08}. The hopping
terms considered are shown in the inset of Fig.\ref{Fig1}(b).
The values of all band parameters except $\gamma_{1}$ and
$\Delta$, which were determined above were taken from Ref.
\onlinecite{PartoensPRB06}: $\gamma_{0}$ = 3.12 eV,
$\gamma_{1}$ = 0.378 eV , $\gamma_{3}$ = 0.29 eV, $\gamma_{4}$
= 0.12 eV and $\Delta$ = 0.015 eV. Note that they agree well
with the values determined in Ref.\cite{MalardPRB07} using
Raman spectroscopy. The doped charge and the Fermi energy can
be directly determined for any given gate voltage using the
known capacitance of the SiO$_{2}$ layer \cite{CastroPRL07}.
The Kubo formula was used to calculate $G(\omega)$ that was
eventually Gaussian-broadened by 0.02 eV, in order to match the
observed line widths. The reflectivity spectra were computed
based on Fresnel equations using the known optical properties
of the SiO$_{2}$/Si substrate. We begin with a calculation
which assumes that the only effect of applying gate voltage is
to shift the chemical potential and does not include the
gate-induced bandgap.

In panels (a) and (c) of Fig.\ref{Fig3}, the color plots of
experimental and calculated spectra of $\Delta R(\omega,V_{g})$
are represented. One can notice a quite good correspondence
between the energy and the gate voltage dependence of the
strong spectral features. Having found that such an agreement
is present in the raw reflectivity data, we proceed with a
detailed experiment-theory comparison in terms of the optical
conductance (Fig.\ref{Fig3}(b) and (d)). In view of the
mentioned possibility that the extracted conductance curves
contain a spectrally featureless background, here we subtract
from all spectra the same, {\em i.e.} gate-voltage independent,
smooth curve shown as a dashed line in Fig.\ref{Fig2}(b). This
curve is chosen in such a way that the corrected $\mbox{Re
}G(\omega, V_{g}=100 \mbox{ V})$ coincides with the theoretical
values in the regions around 0.2 eV and 0.6 eV, where no sharp
structures are expected.

The assignment of the optical conductance structures to
interband transitions is given in Fig.\ref{Fig3}(d). Apart from
the discussed strong peak structures B and C there is an
onset-like structure A which corresponds to a transition
between the low-energy bands 2 and 3, which has the same origin
as the onset-like structure observed in monolayer graphene
\cite{LiNP08}. The onset frequency is twice the Fermi level
with respect to the Dirac energy, which is in bilayer graphene
proportional to $|V_{g}-V_{g0}|$ with a coefficient determined
by $\gamma_{0}$. In the measured spectra (Fig.\ref{Fig3}(b)) we
observe such a structure showing the same (within the
experimental uncertainty) dependence on the gate voltage. This
confirms that $\gamma_{0}$ is close the value used in the
calculation (3.12 eV). This observation is in accordance with a
recent measurement of Li et al. \cite{LiCM08}. Interestingly,
there is a {\em second} onset-like structure, with the onset
energy showing a similar V-shape dependence on the gate voltage
but shifted with respect to the structure A by about
$\gamma_{1}$. The structure is due to the onset of transition D
($1\rightarrow3$) for the electron doping and transition E
($2\rightarrow4$) for the hole doping. There is a significant
enhancement of $\mbox{Re }G(\omega)$ close to the 'vertex'
point $\omega \approx \gamma_{1}$, $V \approx V_{g0}$ where the
two onsets are close to each other \cite{NilssonPRL06,AbergelFalkoPRB07}. One
can clearly see a similar structure on the experimental graph.
Thus the tight binding model reproduces most of the features of
experimental spectra.

Now we address the issue of the gate-induce bandgap
$\Delta_{g}$ between the low-energy electron and hole bands
\cite{McCannFalkoPRL06,GuineaPRB06,McCannPRB06}. Its
manifestation in the infrared spectra was first calculated
(assuming that $\gamma_{3}$, $\gamma_{4}$ and $\Delta $ = 0) in
Ref.\cite{NicolCarbottePRB08}. In Fig.\ref{Fig3}(f) we show the
result of a calculation where we keep the all aforementioned
band parameters and add a gate-dependent difference in
electrostatic potential between the two planes. We use a curve
$\Delta_{g}(V_{g})$ from Ref.\onlinecite{CastroPRL07}, shown as
a red line in Fig.\ref{Fig3}(f), where the charge screening
effects were treated self-consistently. We assume, as it was
also done in Ref.\onlinecite{CastroPRL07}, that contaminant
molecules shifting the charge neutrality point away from $V_{g}
= 0$ act as an effective top gate electrode. In this case the
bandgap vanishes not at $V_{g} = V_{g0}$ but at $V_{g} =
-V_{g0}$. At the highest gate voltages of our experiment the
gap value is expected to be of the order of 0.1 eV.

According to the calculation, the opening of the bandgap indeed
brings some extra features to the spectra. All of them are due
to the flattening of bands 2 and 3, as shown in
Fig.\ref{Fig3}(e), which results in a strong increase of the
density of states of these bands. The first feature (marked A')
is an enhancement of the optical intensity of the transition $2
\rightarrow 3$. Although this enhancement largely shows up at
photon energies below the experimentally accessible region, its
tail spreads up to about 0.2 eV. The second feature is the
appearance of high-frequency satellites (marked E' and D') to
the peak-like structures B and C. These satellites correspond
from transitions $2 \rightarrow 4$ and $1\rightarrow 3$
respectively. The energy separation between the central
frequencies of peaks B and E' as well as between C and D' is
close to the energy of the bandgap and could be therefore read
directly from the conductance curves. Note that the interband
structures A', E' and D' involve the same band pairs as the
structures A, E and D respectively. However the former ones are
exclusively due to transitions within a very small momentum
region around the Dirac point.

We notice that experimental spectra (Fig.\ref{Fig3}(b)) show an
enhancement of conductance similar to the high-frequency tail
of the structure A'. However the satellite structure E' and D'
are not obviously present in the data. Thus the tight binding
model that is quite successful in describing the main infrared
features, is only in partial agreement with the data as far as
the bandgap related features are concerned. This fact is
perhaps the largest surprise of our study. We can only
speculate about the possible reasons. First of all, the
satellite features might be smeared out by doping
inhomogeneity, due to the flake corrugation, contaminant
molecules or other factors. However, the calculation already
takes a large broadening (about 0.02 eV) into account. A second
possibility is that the actual bandgap is smaller than the
prediction of a simple model that does not take into account
interaction effects, so that the satellites E' and D' cannot be
easily separated from the main peaks. A third possibility is that the gap
can be partially filled with impurity states \cite{NilssonPRL07}. Future experimental and
theoretical developments are required to resolve this
intriguing issue.

This work was supported by the Swiss National Science
Foundation through the National Center of Competence in
Research "Materials with Novel Electronic Properties-MaNEP". We
are grateful to A. Morpurgo, L. Benfatto, E. Cappelluti and M.
Fogler for helpful discussions.


\begin{references}

\bibitem{NovoselovScience04}
K.S. Novoselov, A.K. Geim, S.V. Morozov, D. Jiang, Y. Zhang,
S.V. Dubonos, I.V. Grigorieva, A.A. Firsov, Science
\textbf{306}, 666 (2004).

\bibitem{GeimNovoselovNatMat07}
A. K. Geim, K. S. Novoselov, Nature Mater. \textbf{6}, 183
(2007).

\bibitem{AndoJPSJ02}
T. Ando, Y. Zheng, H. Suzuura, J. Phys. Soc. Japan \textbf{71},
1318 (2002).

\bibitem{NairScience08} R.R. Nair, P. Blake, A.N. Grigorenko, K.S.
Novoselov, T.J. Booth, T. Stauber, N.M.R. Peres, A.K. Geim,
Science \textbf{320}, 1308 (2008).

\bibitem{LiNP08}
Z.Q. Li, E.A. Henriksen, Z. Jiang, Z. Hao, M.C. Martin, P. Kim,
H.L. Stormer, and D.N. Basov , Nature Physics \textbf{4}, 532
(2008) .

\bibitem{KuzmenkoPRL08} A. B. Kuzmenko, E. van Heumen, F. Carbone,
D. van der Marel, Phys. Rev. Lett. \textbf{100}, 117401 (2008).

\bibitem{NilssonPRL06}
J. Nilsson, A. H. Castro Neto, F. Guinea, and N. M. R. Peres,
Phys. Rev. Lett. \textbf{97}, 266801 (2006).

\bibitem{AbergelFalkoPRB07}
D. S. L. Abergel and V. I. Fal'ko, Phys. Rev. B \textbf{75}, 155430 (2007).

\bibitem{NicolCarbottePRB08}
E. J. Nicol and J. P. Carbotte, Phys. Rev. B \textbf{77}, 155409, (2008).

\bibitem{JiangPRL07}
 Z. Jiang, E.A. Henriksen, L.C. Tung, Y.-J. Wang, M.E. Schwartz, M.Y. Han, P. Kim, H.L. Stormer,
 Phys. Rev. Lett. \textbf{98}, 197403 (2007).

\bibitem{WangScience08}
F. Wang, Y. Zhang, C. Tian, C. Girit, A. Zettl, M. Crommie, Y.
R. Shen, Science \textbf{320}, 206 (2008).

\bibitem{SWMcC}
J.W. McClure, Phys. Rev. \textbf{108}, 612 (1957); J.C.
Slonzcewski and P.R. Weiss, Phys. Rev. \textbf{109}, 272
(1958).

\bibitem{McCannFalkoPRL06}
D. McCann, V.I. Falko, Phys. Rev. Lett. \textbf{96}, 086805 (2006).

\bibitem{GuineaPRB06}
F. Guinea, A.H. Castro Neto and N.M.R. Peres, Phys. Rev. B \textbf{73}, 245426 (2006).

\bibitem{McCannPRB06}
E. McCann, Phys. Rev. B \textbf{74}, 161403(R) (2006).

\bibitem{OhtaScience06}
T. Ohta, A. Bostwick, T. Seyller. K. Horn, E. Rotenberg,
Science \textbf{313}, 951 (2006).

\bibitem{CastroPRL07}
E. V. Castro et al., Phys. Rev. Lett. 99, 216802 (2007).

\bibitem{OostingaNatMat08}
J. B. Oostinga, H.B. Heersche, X. Liu, A.F. Morpurgo, L.M.K.
Vandersypen, Nature Mater. \textbf{7}, 151 (2008).

\bibitem{KuzmenkoRSI05}
A.B. Kuzmenko, Rev. Sci. Instrum. \textbf{76}, 083108 (2005).

\bibitem{ChungReview}
D. D. L. Chung, J. Mat. Science \textbf{37}, 1 (2002).

\bibitem{GuizzettiPRL73}
G. Guizzetti, L. Nosenzo, E. Reguzzoni, and G. Samoggia, Phys.
Rev. Lett. \textbf{31}, 154 (1973).

\bibitem{LiCM08}
Z. Li et al., cond-mat/0807.3776; L. M. Zhang et al.,
cond-mat/0809.1898.

\bibitem{PartoensPRB06}
B. Partoens and F.M. Peeters, Phys. Rev. B {\bf 74}, 075404
(2006).

\bibitem{MalardPRB07}
L.M. Malard, J. Nilsson, D.C. Elias, J.C. Brant, F. Plentz,
E.S. Alves, A. H. Castro Neto, and M. A. Pimenta, Phys. Rev. B
\textbf{76}, 201401(R) (2007).

\bibitem{NilssonPRL07}
J. Nilsson and A.H. Castro Neto, Phys. Rev. Lett. 98, 126801 (2007).

\end{references}
\end{document}